\documentclass{aa}
\usepackage{graphicx}
\begin{document}
\newcommand{\kms}{{\rm km~s}^{-1}}
\newcommand{\ho}{{\rm km~s}^{-1}~{\rm Mpc}^{-1}}
\newcommand{\sbs}{SBS~1520+530}
\title{Time delay and lens redshift for the doubly imaged BAL quasar \sbs
\thanks{Based on observations made with the Nordic Optical, operated 
on the island of La Palma jointly by Denmark, Finland, Iceland, Norway, 
and Sweden, in the Spanish Observatorio del Roque de los Muchachos of 
the Instituto de Astrofisica de Canarias and the W. M. Keck Observatory
which is operated jointly by the California Institute of Technology,
the University of California and the National Aeronautics and Space
Administration.}}


\author{I. Burud \inst{1}
\and J. Hjorth \inst{2}
\and F. Courbin \inst{3,4}
\and J. G. Cohen \inst{5}
\and P. Magain \inst{3}
\and A. O. Jaunsen \inst{6}
\and A. A. Kaas \inst{7}
\and C. Faure \inst{6,8}
\and G. Letawe \inst{3}
          }

 \offprints{burud@stsci.edu}

\institute{Space Telescope Science Institute, 3700 San Martin Drive,
 Baltimore, MD 21218, USA
 \and
 Astronomical Observatory, University of Copenhagen, Juliane
  Maries Vej 30, DK--2100 Copenhagen \O, Denmark
 \and 
  Institut d'Astrophysique et de G\'eophysique, Universit\'e de
 Li\`ege, All\'ee du 6 ao\^ut, Bat B5C, Li\`ege 1, Belgium
 \and 
 Universidad Cat\'olica de Chile, Departamento de 
 Astronomia y Astrofisica,
 Casilla 306, Santiago 22, Chile
 \and
 California Institute of Technology, Pasadena, CA 91125
 \and
 European Sourthern Observatory, Casilla 19, Santiago, Chile
 \and
 Nordic Optical Telescope, Apartado 474, E-38700 Santa
             Cruz de La Palma, Canary Islands, Spain
\and
Observatoire Midi-Pyr\'en\'ees, 14 avenue Edouard Belin, 31400 Toulouse, 
France 
}

   \date{}

\abstract{We present optical $R$-band light curves of the
gravitationally lensed quasar \object{SBS1520+530} derived from data obtained at
the Nordic Optical Telescope.  A time
delay of $130\pm3$ days (1$\sigma$) is determined from the light curves. In
addition, spectra of \object{SBS1520+530} obtained at the Keck Observatory 
are spatially deconvolved in order
to extract the spectrum of the faint lensing galaxy, free of any
contamination by the light from the bright quasar images. 
This spectrum indicates a lens redshift $z=0.717$, in agreement with one of 
the absorption systems found in the quasar spectra. 
The best mass model of the system includes
a second nearby galaxy and a cluster of galaxies in addition to
the main lensing galaxy. Adopting this model and an $\Omega=0.3$,
$\Lambda=0.7$ cosmology, our
time-delay measurement yields a Hubble constant of H$_{0}=51 \pm 9\,
{\rm km}~{\rm s^{-1}}~{\rm Mpc^{-1}}$ (1$\sigma$ error).
}
\maketitle

\section{Introduction}

Prompted by the successful optical measurements of time delays in
PG~1115+080 (Schechter et al. \cite{Sche}), a photometric monitoring
campaign of gravitationally lensed quasars has been carried out at the
Nordic Optical Telescope (NOT) with the aim of measuring time delays
for many gravitationally lensed quasars.  This observable is of
crucial importance to study the mass distribution in lensing galaxies
and to determine $H_{0}$ using the method proposed by Refsdal (1964).

Two time  delays have  been measured so  far at  the NOT: one  for the
doubly imaged quasar B~1600+434, lensed by an edge-on spiral (Burud et
al. \cite{Burud}) and another for the two (summed) components 
RX~J0911+0550 (Hjorth et al.
\cite{Hjorth}).  In  this paper we  will present our third  time delay
measurement: that of  \object{SBS1520+530}, a  doubly imaged  BAL
quasar  at   $z=1.86$.   \sbs\,   was  discovered  by   Chavushyan  et
al. (\cite{Chavushyan}) as a  double quasar with angular separation of
1.56\arcsec.   The   lensing  galaxy  was  detected   by  Crampton  et
al.(\cite{Crampton}) on $H$-band images obtained by using the adaptive
optics  system  of the  Canada-France-Hawaii  Telescope.  The  lensing
galaxy is also well resolved on optical images obtained at the NOT and
deconvolved using the MCS  deconvolution algorithm (Magain, Courbin \&
Sohy \cite{Magain}), and on public Hubble Space Telescope near-IR data
(Faure  et al.  \cite{Faure}).  The redshift  of  the galaxy  however,
remains  unknown. With  the aim  of measuring  this redshift,  we have
obtained a spectrum of \sbs\, with ESI at the Keck observatory.

The observations  and data  reduction of the  images are  presented in
Sect.~\ref{sect:data} and \ref{sect:phot}. The  time delay measurement is
described in Sect.~\ref{sect:timedelay} while the analysis  of the spectroscopic
data  is explained  in  Sect.~\ref{sect:sbsspec}. This  section  also
includes  a discussion  on the  spectral differences  between  the two
quasar images.   Mass  models  and  the estimate  of  the  Hubble
constant    are    discussed    in    Sect.~\ref{sect:h0}.     Finally,
Sect.~\ref{sect:discussion} summarises the main results.


\section{Observations and Data Reduction}
\label{sect:data}

Weekly  observations  of \sbs\,  were  carried  out  at the  NOT  from
February 1999  to May 2001.  The detector ALFOSC (AndaLuc\'\i  a Faint
Object SpeCtrograph and Camera) with a pixel size of 0\farcs188 was used for 95\%
of the  frames obtained.  The remaining 5\%  were obtained  with HiRAC
(High Resolution Adaptive Camera)  or with the stand-by camera StanCam
which have pixel sizes of 0\farcs107 and 0\farcs176 respectively.

The target  is relatively bright  (I$\sim$18) and an exposure  time of
300 sec  in the $R$-band was  sufficient to obtain  a S/N$\sim$100 for
each quasar image.  The seeing varied from 0\farcs5 to 2\farcs2 during
the two years of observations, 1\farcs0 being the most frequent value.

Our  automated  pipeline employing  routines  in the  IRAF/NOAO/CCDRED
package  was  used  in order  to  pre-process  the  CCD frames  in  an
efficient and homogeneous way.


\begin{figure}   
\centering
\includegraphics[width=8.7cm]{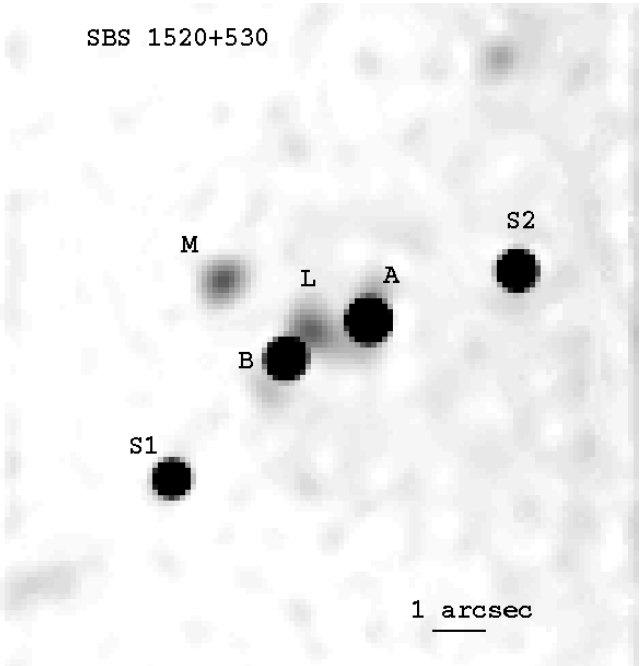}
\caption{Deconvolved $R$-band  image (FWHM$=$0\farcs28) of 
12\arcsec  x 12\arcsec  around
\object{SBS1520+530} obtained from  the simultaneous
deconvolution of 59 frames.   We see  the two stars just to the North-West
(S2)  and South-East (S1), the
quasar  components A  and B,  the  lensing galaxy L close to  B and  an
additional galaxy M, NE of  the lens.  North  is up  and East is  to the
left.} 
\label{ima1} 
\end{figure}%

\begin{figure}
\centering
\includegraphics[width=8.7cm]{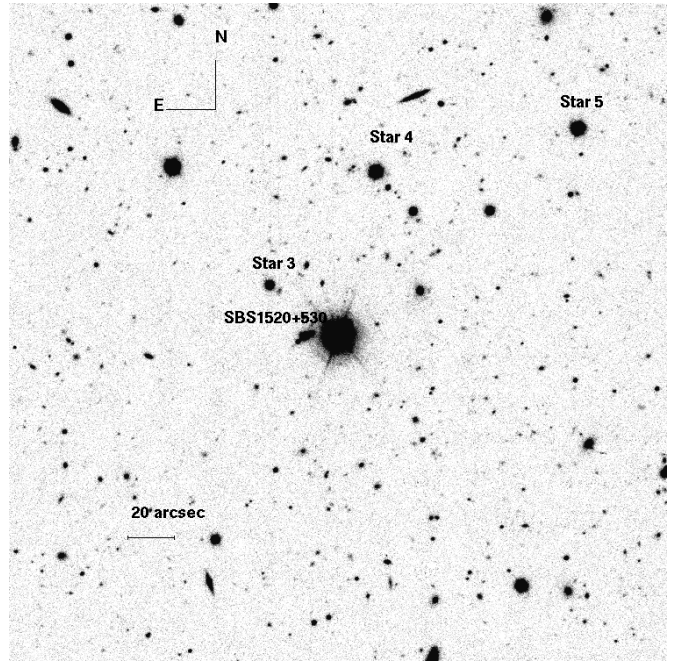}
\caption{The     field     ($\sim$5\arcmin$\times$5\arcmin)     around
\object{SBS1520+530} observed with ALFOSC at the NOT.  North is up and
East to the left. See Faure et al. \cite{Faure} for a detailed 
investigation of the spatial distribution of secondary lenses in this field}
\label{ima2}
\end{figure}


\section{Photometry}
\label{sect:phot}

The photometry of the blended  quasar images was performed by applying
the    MCS    deconvolution     algorithm    (Magain,    Courbin    \&
Sohy~\cite{Magain}).   This algorithm  has  already been  successfully
applied to several  monitoring data of lensed quasars  (e.g., Burud et
al.~\cite{Burud}, Burud et  al.~\cite{Burudb}).  Its main advantage is
its  ability  to  deconvolve  all  the frames  from  different  epochs
simultaneously, hence  constraining the  astrometry of the  two quasar
images and the shape of the  lensing galaxy using the total S/N of the
whole data-set.   The intensity  of the point  sources are  allowed to
vary from image to image, which produces the light curves.

The two  quasar components are  clearly distinguished on  our deconvolved
image of  \object{SBS1520+530} (see Fig.~\ref{ima1}),  and the lensing
galaxy  is resolved.  The lensing  galaxy can  easily be
taken into account  in order to avoid contamination of  the photometry  of the  quasar
images.

\begin{figure}
\centering
\resizebox{\hsize}{!}{\includegraphics{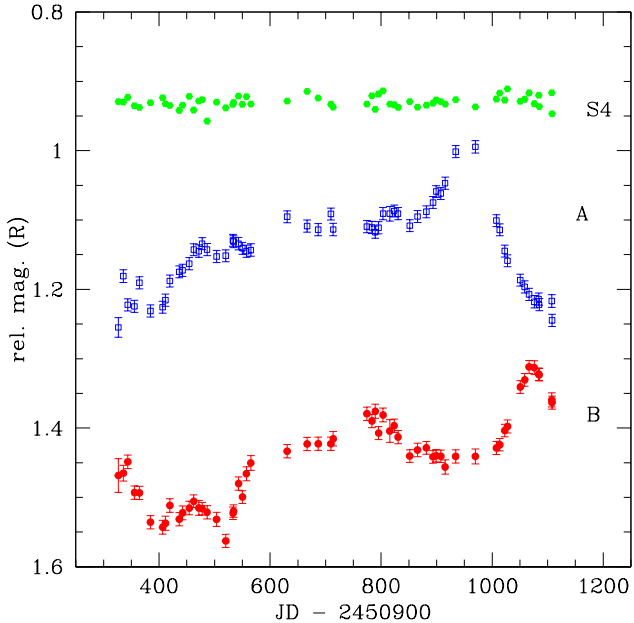}}
\caption{$R$-band lightcurves of  \object{SBS1520+530} and of the
reference  star S4  (see Fig.~\ref{ima2}),  calculated relative  to the
reference  star S3  (Fig.~\ref{ima2}).  The  error bars  include photon
noise and PSF  errors estimated from the deconvolution  of a reference
star.  The B component is shifted vertically by $-0.4$ mag and star S4 is 
shifted by $+0.5$ mag on the plot.}
\label{lightcurves}
\end{figure}

The light curves of \object{SBS1520+530}  consist of 57 data points in
the $R$-band  as presented in Fig.~\ref{lightcurves} and Table~\ref{data}.   
In addition to
the two  stars just to the South-East (S1) and  North-West (S2) 
of the quasar (Fig.~\ref{ima1}), three
other reference stars in the field were deconvolved, in order to check
the  relative  photometry and  to  check  for  systematic errors 
(stars 3 to 5 in
Fig.~\ref{ima2}).   The magnitudes of the quasar  are relative  to  
the magnitudes of Star  3  on
Fig.~\ref{lightcurves}.   The  error-bars   include  both  photon  noise  and
additional  systematic  errors,  e.g.,   PSF  errors.  The  latter  is
estimated by using the reference  stars, as explained in 
Burud et al. (\cite{Burud}).

\section{Time delay measurement}
\label{sect:timedelay}

\begin{figure*}
\centering
\includegraphics[width=5.5cm]{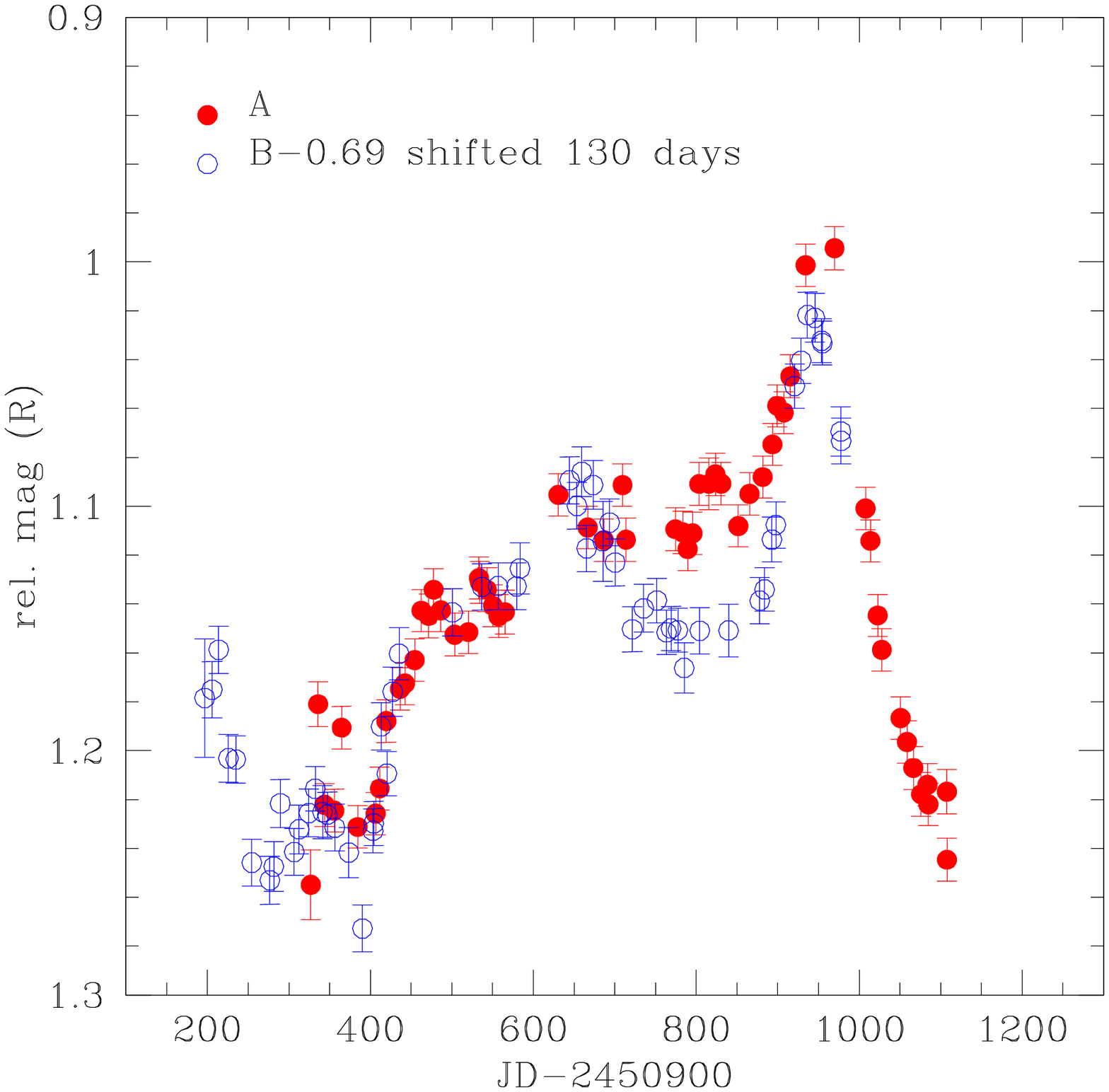}
\hspace{1mm}
\includegraphics[width=5.5cm]{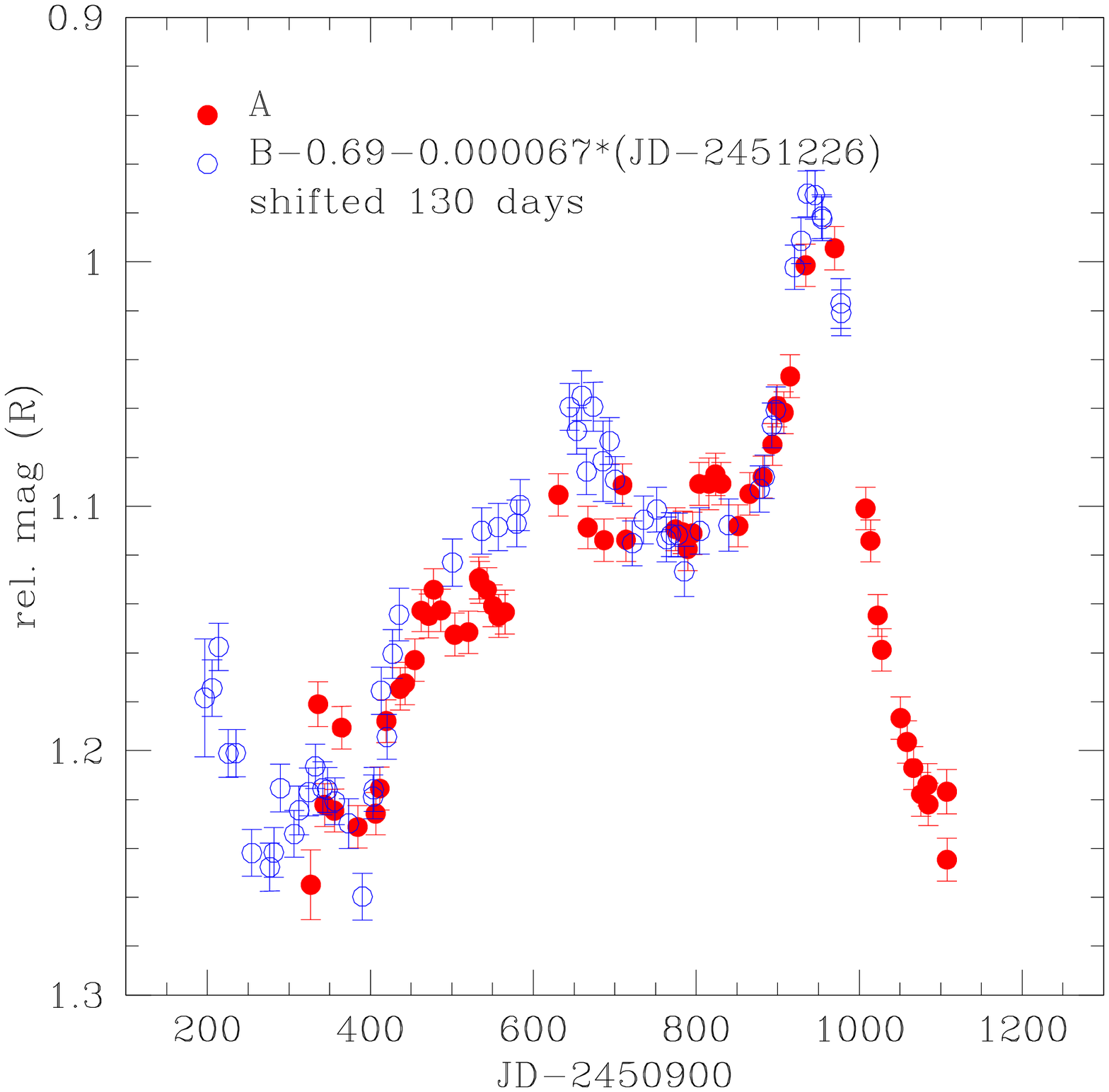}
\hspace{1mm}
\includegraphics[width=5.5cm]{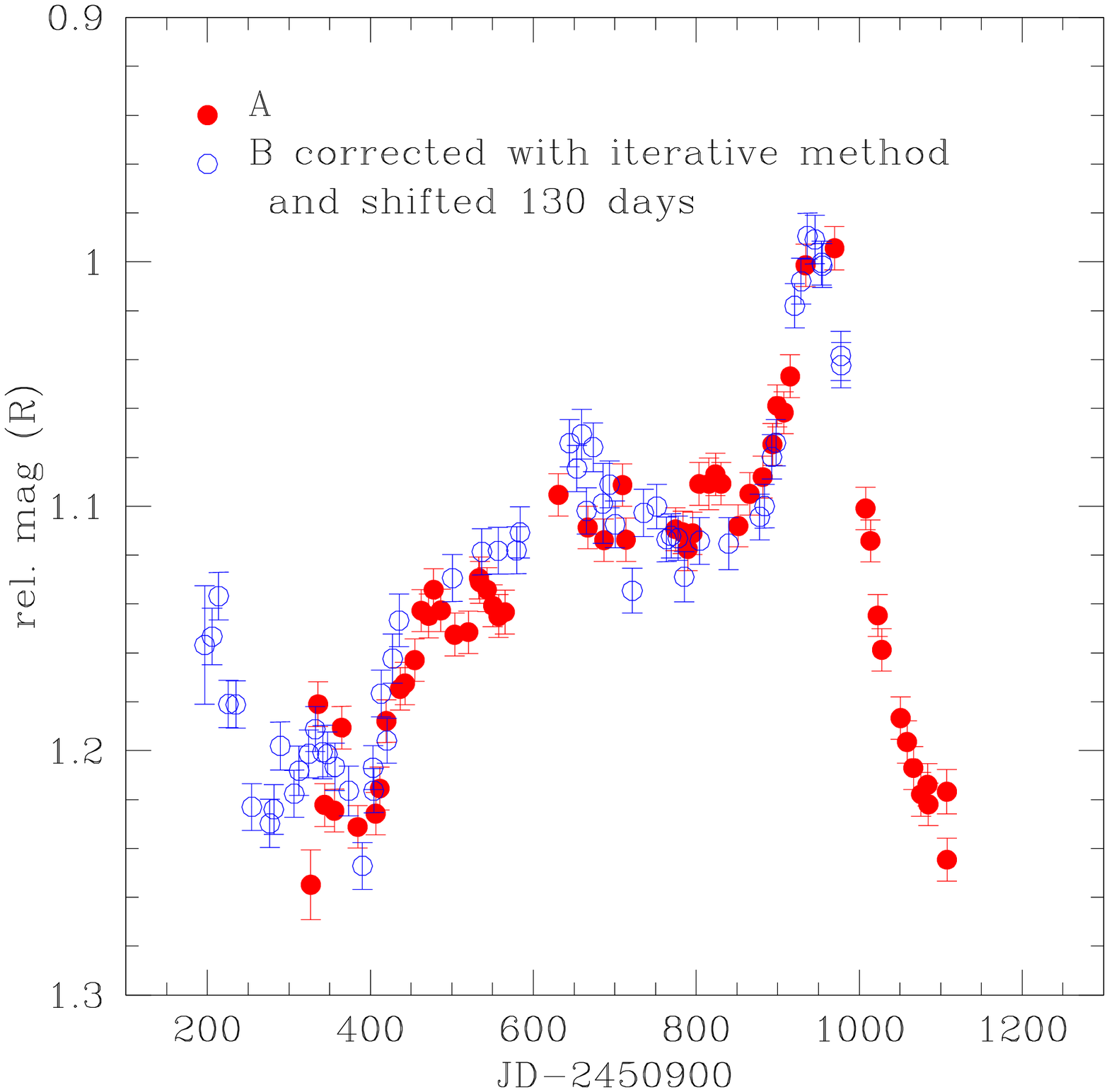}
\caption{Time  delay  shifted  light curves  of  \object{SBS1520+530}.
{\it  Left}: B  is shifted  by 0.69  mag, no  correction  for external
variations (e.g., as can be  induced by microlensing) is applied. {\it
Middle}: B  is shifted  by 0.69  mag and a  linear correction  for the
external variations is applied. {\it  Right}: B is shifted vertically by 
0.69 mag
and  corrected  for  external  variations with  the  iterative  method
(cf. Sect.~\ref{sect:timedelay}). }
\label{shift}
\end{figure*}

By sliding  the light  curves across  one another one  can derive  a 
rough  time  delay  estimate  of  $\sim$ 125  days.  As  predicted  by
gravitational lens theory, the brightest quasar image A is the leading
component.  Using the $\chi^2$  minimization method described in Burud
et al. (\cite{Burudc}),  a time delay value of  $128 \pm 3$ (1$\sigma$)
days  is  found  from   the  $R$-band  light  curves.   The  magnitude
difference between the A and  the B component is $0.692\pm0.003$ mag,
corresponding to an  A/B flux ratio of $1.892\pm0.002$.   The error bar
on the  time delay  is obtained from  Monte Carlo simulations  of 1000
sets  of  light  curves,  assuming  that the  photometric  errors  are
uncorrelated  and   follow   a  Gaussian  distribution.   The  best
$\chi^2$-fit   is  obtained   when  a   linear  term,   modelling  slow
microlensing effects, is added to one of the components. The slope of
this  term  is  determined  by  the  algorithm.   With  this  external
variation, the shifted  light  curve  of the  B  component plotted  in
Fig.~\ref{shift} becomes $B'=B-0.692-0.000067 \times(JD-2451226)$.  The
linear correction does not  remove all the external variations. Faster
variations on  time scales  of $50-100$ days  are still  present.  The
iterative version  of the algorithm  (Burud et al.  \cite{Burudc}) was
also  applied in  an attempt  to correct for these fast  variations.  This
method yields  $\Delta t  = 130\pm3$ days,  slightly higher  than the
value found with  the direct method but still  in agreement within the
error  bars.  Even  with  the  iterative method  some  of the  fastest
variations are not corrected for (Fig.~\ref{shift}), notably the small
peak in the A component at $\sim$JD 2451750 (abscissa 850 in Fig.~\ref{shift}).


\section{Spectroscopy of SBS1520+530}
\label{sect:sbsspec}

\subsection{Keck II spectroscopy}

Converting the  time delay  into an estimate  of the  Hubble parameter
requires  good  knowledge  of  the  geometry of  the  lensing  system,
including the redshift  of the source and of  the lensing galaxy.  For
this purpose, we obtained a  deep optical spectrum of SBS1520+530 with
the Keck II Telescope and the Echellette Spectrograph and Imager (ESI)
(Sheinis et al. \cite{sheinis}).
The  observations were  obtained on  May  23, 2001  under good  seeing
conditions (0.6\arcsec).   The 1\arcsec\,  slit was aligned  to obtain
simultaneously the  spectrum of SBS1520+530 and of  star S2 (indicated
on Fig.~\ref{ima1}).   Four 900s exposures  were taken so  that cosmic
rays could be efficiently removed.

\begin{figure}
\centering
\resizebox{\hsize}{!}{\includegraphics{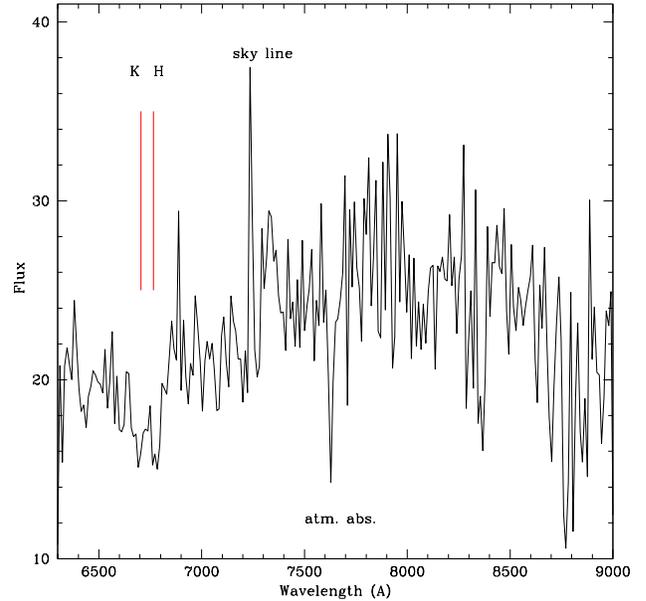}}
\caption{Spectrum  of the  lensing  galaxy in  SBS1520+530.  The  CaII
($\lambda$ 3933, 3968 {\AA}) is  identified.  Also indicated are a 
night sky
emission line and an atmospheric absorption line.}
\label{galaxy}
\end{figure}

\begin{figure}
\centering
\resizebox{\hsize}{!}{\includegraphics{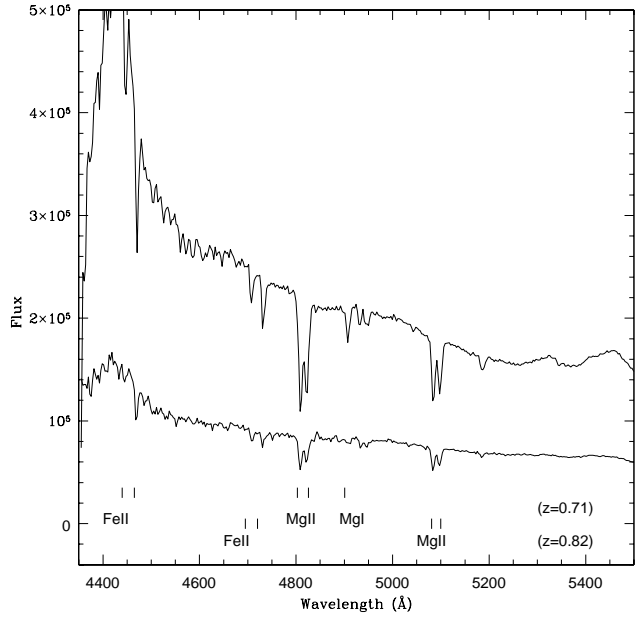}}
\resizebox{\hsize}{!}{\includegraphics{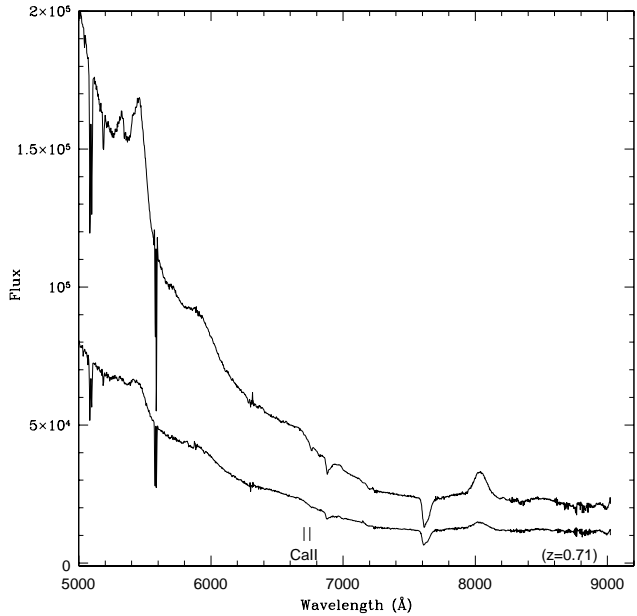}}
\caption{Spectra of the quasar components A (top) and B (bottom) of
SBS~1520+530. Absorption lines corresponding to metal absorptions
at $z=0.71$ and $z=0.82$ are indicated.}
\label{specAB}
\end{figure}

\begin{figure}
\centering
\resizebox{\hsize}{!}{\includegraphics{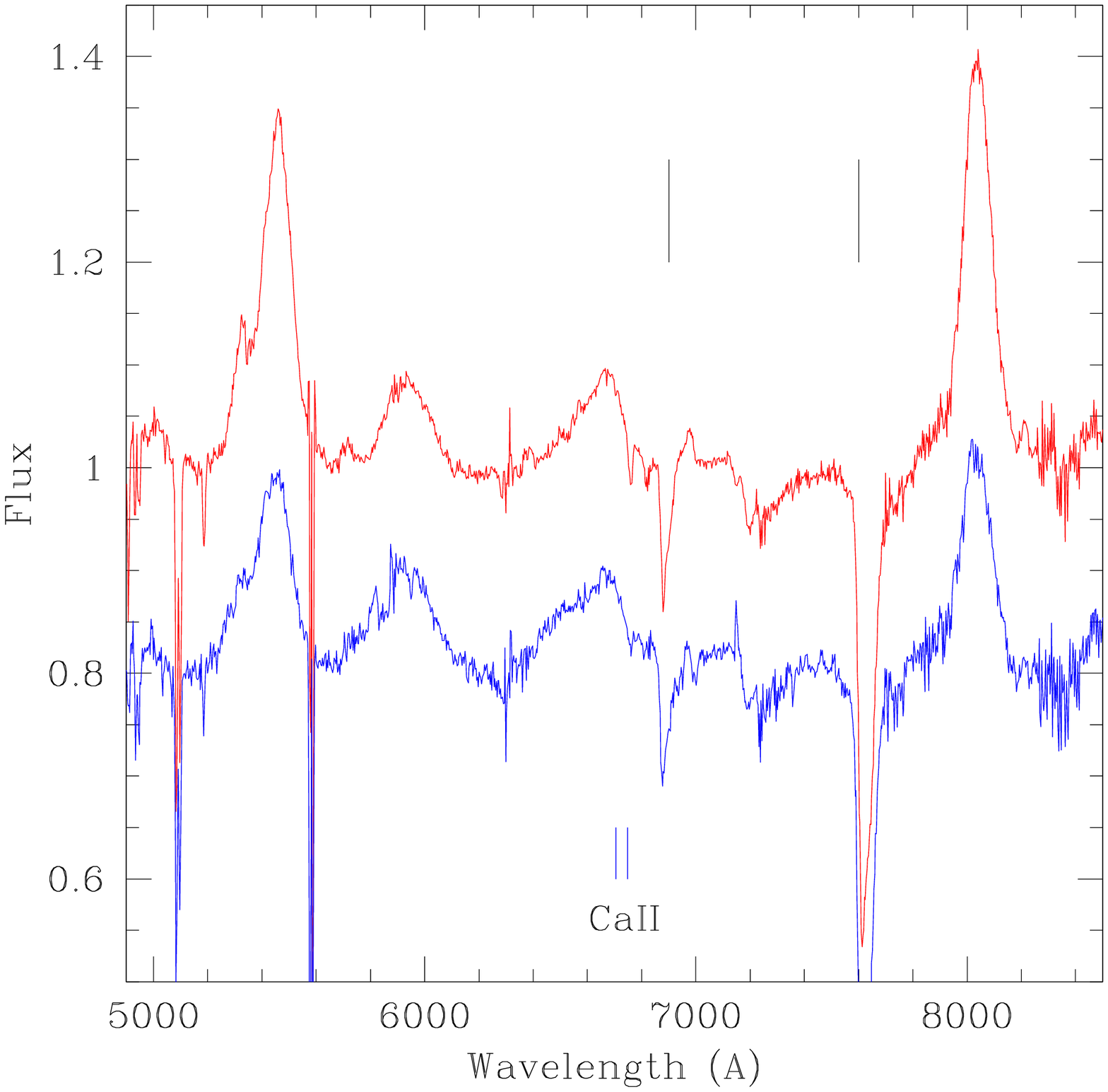}}
\caption{The A (top) and B (bottom, vertically shifted by -0.2
for clarity) spectra with normalized 
continuum. The equivalent widths of the emission lines
including AlIII/CIII] and MgII are significantly smaller  in B
than in A. The CaII absorption lines at redshift 0.71 are 
indicated as well as two telluric bands.}
\label{norm}
\end{figure}

\subsection{Redshift of the lensing galaxy}

Using the  spatial information  contained in the  spectrum of  the PSF
star  S2, the  2D spectrum  of  \sbs\, was  spatially deconvolved,  by
applying  the  spectral version  of  the  MCS deconvolution  algorithm
(Courbin  et al.,  2000b).  The  deconvolution process  decomposes the
data into the  individual spectra of the two quasar  images and of the
faint  lensing galaxy (see  Figs.~\ref{galaxy} and  ~\ref{specAB}), as
done for  the lensed systems HE~2149$-$2745 (Burud  et al. \cite{Burudb})
or HE~1104$-$1805 (Lidman et al. \cite{lidman}).  No emission lines were
detected in  the spectrum  of the lensing  galaxy (Fig.~\ref{galaxy}).
However, an  absorption doublet is  clearly observed at  $\lambda \sim
6700$ {\AA}, matching within the error-bars  the CaII H and K lines at
redshift 0.71.   This strongly supports  a scenario where  the lensing
galaxy  is  responsible for  the  CaII  absorption  lines at  redshift
$z=0.71\pm0.005$ first  found by Chavushyan  et al. (\cite{Chavushyan})
in  the  quasar  spectra.   Another  absorption system  was  found  by
Chavushyan et  al. (\cite{Chavushyan}), at $z=0.82$. This  system does
not seem to be associated with the lens.

\subsection{Spectra of the quasar components}

The quasar spectra display the usual prominent emission lines such as
AlIII/CIII] ($\lambda$ 1909 {\AA}) and MgII ($\lambda$ 2798 {\AA}), at
redshift $z=1.855$.  Both (deconvolved) spectra show the metal
absorption lines first reported by Chavushyan et al.  (1997).  These
are the FeII ($\lambda$ 2587, 2600 {\AA}), MgII ($\lambda$ 2798{\AA}),
MgI ($\lambda$2853 {\AA}) absorption lines at redshifts
$0.717\pm0.001$ and $0.817\pm0.001$. CaII ($\lambda$ 3933, 3968 {\AA})
is seen only at $z=0.717$ and is indicated on Fig.~\ref{specAB}.  It
is better seen in the normalized spectra of Fig.~\ref{norm}.

The general shape of the  spectra towards the blue wavelengths is most
probably corrupted by  atmospheric refraction.  The observed reddening
of B compared to A at  the blue wavelengths may therefore 
not be real. 
In the  red parts of  the spectra,  we find a  flux ratio of  1.82 (at
$\lambda$ 9000 {\AA}) which, compared with the ratio of 1.892 measured
on $R$ band  light curves ($\lambda$ 6500 {\AA}),  indicates an almost
constant flux ratio in this wavelength range.

The  continuum normalized  spectra, displayed  in  Fig.~\ref{norm}, show
that the equivalent widths of the  emission lines are much larger in A
than  in  B, suggesting  that  B's  continuum  is amplified.   Such  a
differential amplification  can be explained  by microlensing effects.
Assuming that the lensed quasar follows the ``standard'' AGN model and
that  we are  observing  component  B through  a  network of  caustics
produced by stars in the main lensing galaxy, one can imagine that the
inner continuum region  of component B is enhanced  by a larger amount
than the  much larger emission  line region.  This  interpretation has
already been proposed to explain similar spectral differences observed
in  the double  HE~1104$-$1805 (Wisotzki  et al.  \cite{wis},  Courbin et
al. \cite{Courbin_b}) and HE~2149$-$2745 (Burud et al. \cite{Burudb}).

\section{Modelling the lens}
\label{sect:h0}

The time delay measured for \sbs\, can be used to infer an estimate of
$H_0$, via  a model of  the total gravitational  potential responsible
for the  lensing effect. This model  shall take into  account the main
lensing galaxy and a small galaxy cluster located to the North-West of
the quasar.

Our models, already described  in Faure et al.  (\cite{Faure}), assume
that  different   mass  components  along   the  line  of   sight  are
Pseu\-do-Iso\-ther\-mal-El\-lipti\-cal-Mass-Dis\-tribu\-tion    (PIEMD,
Kassiola \& Kovner \cite{Kass}, Kneib et al.  \cite{Kneib93}, Kneib et
al.   \cite{Kneib96}).  In addition  to the  observational constraints
already available in  Faure et al. (\cite{Faure}), we  now also try to
pin down the flux ratio between  the two quasar components, and to use
it in  the models. This  ratio is difficult  to estimate for  \sbs: it
varies with wavelength,  it has different values in  the continuum and
in  the emission  lines, and  even varies  within the  emission lines.
Microlensing shall  be blamed, but one  should also keep  in mind that
our spectra  were not  taken at parallactic  angle, with  a relatively
narrow slit, and at large  airmass. The values we derive are therefore
only tentative: m$_B$ - m$_A$ = 1.40~mag (i.e.,  $F_A/F_B=3.6$) in the
emission lines, and  m$_B$ - m$_A$ = 0.83~mag (i.e., $F_A/F_B=2.15$) in
the  continuum   The  latter   value  was measured from $K$-band images  
(Faure  et al. \cite{Faure}). 

Table  \ref{tabtab} presents two lens  models, computed  for  the two
 flux  ratios,   and  matching   a   time-delay  of
130$\pm$3~days.  Model  (L) considers only  the main lensing  galaxy L
(see \ref{ima1})  with a measured  ellipticity $\epsilon =0.50 \pm 0.2$
and with  a position  angle $PA=-23 \deg \pm  7\deg$ (see Faure  et al.
2002).  Model (L+M+C) includes the  main lensing galaxy (L), galaxy M,
located  at  2.6\arcsec~  North-East  from  the quasar  image  A  (see
Fig.~\ref{ima1}), and a galaxy cluster, centered at $\sim$1\arcmin~ to
the North-West of the quasar.

\begin{table*}[t]
\begin{center}
\caption[]{Lens models: the first column describes the lens parameters,
where   $\gamma$ and $\theta_{\gamma}$ are the intensity and orientation
of the shear at the position of the quasar images. Column (2) and (3)
gives the models results for the two possible flux ratio measured
in \sbs. The letter $\sigma$ 
refers to lens velocity dispersions and r$_{cut}$ is the cut radius
of the different mass distributions.
\label{tabtab}}
\begin{tabular}{|l|c|c|}
\hline
  M$_B$-M$_A$  &0.83 mag  &1.4 mag   \\ 
\hline
\hline 
 &                 & \\
L& no good fit     & $\sigma_L$=228 km~s$^{-1}$ r$_{cut}$(L)=13 kpc   \\
 &                 &  \\
\hline
$\gamma$ &  & 0.39 \\
$\theta_{\gamma}$ &  & $+$75.3 $\deg$ \\
\hline
H$_{0}$ (1$\sigma$) &  & 63 $\pm$ 9~km~s$^{-1}$Mpc$^{-1}$ \\ 
\hline
\hline
 & $\sigma_L$=189 km~s$^{-1}$ r$_{cut}$(L)=51 kpc & $\sigma_L$=200 km~s$^{-1}$ r$_{cut}$(L)=100 kpc \\
L+M+C   & $\sigma_M$=135 km~s$^{-1}$ r$_{cut}$(M)=20 kpc & $\sigma_M$=118 km~s$^{-1}$ r$_{cut}$(M)=40 kpc \\
   & $\sigma_C$=718 km~s$^{-1}$ r$_{cut}$(C)=600 kpc & $\sigma_C$=400 km~s$^{-1}$ r$_{cut}$(C)=600 kpc \\
\hline
$\gamma$ &  0.32  & 0.34 \\
$\theta_{\gamma}$ & $+$74.5 $\deg$ & $+$74.3 $\deg$ \\
 \hline
H$_{0}$ (1$\sigma)$ & 52 $\pm$ 8 ~km~s$^{-1}$Mpc$^{-1}$ & 50 $\pm$ 8 ~km~s$^{-1}$Mpc$^{-1}$\\ 
\hline
\end{tabular}
\end{center}
\end{table*}


Modeling \sbs\, is possible with a single lensing galaxy (i.e., galaxy
L), only if  m$_B$ - m$_A$ $>$ 1.15. Given the errors  on the image flux
ratio, such a model is  plausible. Taking the emission line flux ratio
(column 3 in  Table \ref{tabtab}) as the true one,  because it is less
affected   by  microlensing,   we  find   H$_0$  = 63$ \pm 9~\rm
km~s^{-1}~Mpc^{-1}$. The  error, once the  image flux ratio  is fixed,
comes mainly from the position  angle of the lens and its ellipticity.

If  the magnitude  difference between  the images  is less  than 1.15,
\sbs\, can  not be modeled  by using one  single galaxy. The  model in
column (2)  of Table  \ref{tabtab}, with galaxy  M and  an intervening
galaxy cluster gives 52 $\pm~ 8~ \rm km~s^{-1}~Mpc^{-1}$.

Finally, changing the flux ratio does not change much
the value  of H$_0$,  for a  given lens model.  The flux ratio is
important in the  choice  of model  (i.e.,
multiple or single lens) but does not alter much the fit once the model is 
given.

\section{Discussion}
\label{sect:discussion}

The  time delay  $\Delta t  =  130 \pm3$ ~days (1  $\sigma$) has  been
measured for the first time, in  the lensed quasar \sbs\, on the basis
of $R$-band  images obtained with  the NOT.  Keck spectroscopy  of the
lensing  galaxy  strongly  suggests  that  the  absorption  system  at
$z=0.717$  is  associated  with  the  lensing  galaxy.   

Applying the detailed mass model presented by Faure et al.
(\cite{Faure}) we derive a mean value H$_0= 51 \pm9~ \rm km
s^{-1}Mpc^{-1}$. When only the main lensing galaxy is used in the mass
model H$_0$ increases to 63 $\pm 9~ \rm kms^{-1}Mpc{-1}$.  The fit
of the wo different models depends on the flux ratio between the two
images, the model with only one galaxy only gives a good fit when the
high flux ratio measured from the emission lines is applied.
Determining precisely the flux ratio is therefore crucial to further
improve the modeling of the system and hence the precision on H$_0$.
Finally, a possible systematic error adds to the uncertainty in H$_0$ 
if the central mass concentration is not isothermal.
 
External  variations,  probably   due  to  microlensing  effects,  are
observed    on     the    time    delay     shifted    light    curves
(Fig.~\ref{lightcurves}).   Part of this  effect can  be modeled  as a
linear  term,  or  corrected  for  with the  iterative  algorithm  for
measuring  time delays.  However,  significant external  variations of
time scales of $\sim$50 days do remain once these corrections are made.

Microlensing is important in \sbs\, as suggested both by the high frequency
variations in the light curves and by the spectra of the quasar images. 
With light curves in only one band we can not efficiently disentangle 
between microlensing and intrinsic variations of the quasar. Such a 
work shall be possible using the colour information provide by 
a monitoring of the object in spectroscopy. The issue in conducting
such a spectrophotometric monitoring for \sbs\, is double: (1) to measure
the true flux ratio between the quasar images and discriminate
between lens models and (2) to use microlensing to infer constraints
on the distribution of micro-lenses in galaxy (L) and/or to reconstruct
the energy profile of the central AGN in the source.


\begin{acknowledgements}

We thank  the NOT  Director Vilppu Piirola  for granting  us observing
time for this project on  a flexible basis. We are especially grateful
to the  many visiting  observers at NOT  who have contributed  to this
project  by performing  the scheduled  observations. This  project was
conceived in 1997 while JH and AOJ were visiting scientists
at the Center  for Advanced Study in Oslo.  We  thank Rolf Stabell and
Sjur  Refsdal for  inviting  us in  their  center and  for their  kind
hospitality. JH appreciates  the hospitality of the OMP  where some of
this  work was  conducted.  The  project was  supported by  the Danish
Natural Science Research  Council (SNF).  IB was supported  in part by
contract  ARC94/99-178   ``Action  de  Recherche   Concert\'ee  de  la
Communaut\'e   Fran\c{c}aise  (Belgium)''   and   P\^ole  d'Attraction
Interuniversitaire, P4/05  \protect{(SSTC, Belgium)}.  FC acknowledges
financial support  through Chilean grant  FONDECYT/3990024 and through
Marie  Curie  grant   MCFI-2001-0242.   Additional  support  from  the
European Southern Observatory and ECOS/CONICYT CU00U05 
grant is also
gratefully acknowledged.

\end{acknowledgements}

\clearpage

\begin{table}[t]
\begin{center}
\caption[]{R-band photometry of SBS~1520+530} 
\begin{tabular}{lcccccc}
\hline
   A   & $\sigma_{A}$ & B & $\sigma_{B}$ & JD \\  
\hline
  1.2548 & .0143 & 1.868 & .0242 & 2451226.75 \\
  1.1809 & .0091 & 1.865 & .0115 & 2451235.75 \\
  1.2221 & .0087 & 1.848 & .0096 & 2451243.75 \\
  1.2245 & .0087 & 1.893 & .0097 & 2451255.75 \\
  1.1905 & .0086 & 1.893 & .0096 & 2451264.75 \\
  1.2312 & .0086 & 1.935 & .0095 & 2451284.50 \\
  1.2257 & .0087 & 1.943 &.0098  & 2451306.50 \\
  1.2154 & .0087 & 1.937 & .0101 & 2451311.50 \\
  1.1879 & .0087 & 1.911 & .0097 & 2451319.50 \\
  1.1747 & .0086 & 1.931 &.0096  & 2451336.50 \\
  1.1725 & .0087 & 1.922 & .0099 & 2451342.50 \\
  1.1629 & .0087 & 1.915 & .0097 & 2451354.50 \\
  1.1426 & .0085 & 1.905 &.0091  & 2451362.50 \\
  1.1448 & .0089 & 1.915 & .0109 & 2451371.50 \\
  1.1341 & .0086 & 1.916 & .0092 & 2451377.50 \\
  1.1426 & .0086 & 1.921 & .0096 & 2451386.50 \\
  1.1525 & .0087 & 1.931 & .0102 & 2451403.50 \\
  1.1515 & .0086 & 1.963 &.0095  & 2451420.25 \\
  1.1293 & .0085 & 1.922 & .0090 & 2451433.25 \\
  1.1310 & .0085 & 0.920 & .0089 & 2451434.25 \\
  1.1341 & .0089 & 1.880 & .0096 & 2451443.25 \\
  1.1406 & .0086 & 1.899 & .0091 & 2451450.25 \\
  1.1449 & .0087 & 1.865 & .0099 & 2451457.25 \\
  1.1433 & .0089 & 1.850 & .0107 & 2451465.25 \\
  1.0952 & .0087 & 1.833 & .0096 & 2451530.75 \\
  1.1086 & .0086 & 1.822 & .0094 & 2451566.75 \\
  1.1139 & .0087 & 1.822 & .0096 & 2451586.75 \\
  1.0913 & .0086 & 1.822 & .0095 & 2451609.50 \\
  1.1137 & .0089 & 1.815 & .0104 & 2451613.75 \\
  1.1094 & .0086 & 1.779 & .0096 & 2451674.50 \\
  1.1107 & .0086 & 1.789 & .0094 & 2451683.50 \\
  1.1174 & .0088 & 1.775 & .0101 & 2451689.50 \\
  1.1111 & .0086 & 1.807 &.0095  & 2451695.50 \\
  1.0908 & .0087 & 1.781 & .0100 & 2451703.50 \\
  1.0908 & .0105 & 1.804 & .0163 & 2451715.50 \\
  1.0869 & .0086 & 1.796 & .0095 & 2451723.50 \\
  1.0907 & .0086 & 1.813 & .0096 & 2451730.50 \\
  1.1080 & .0086 & 1.840 & .0091 & 2451751.50 \\
  1.0948 & .0087 & 1.831 & .0097 & 2451765.50 \\
  1.0880 & .0085 & 1.828 & .0091 & 2451781.50 \\
  1.0747 & .0086 & 1.842 &.0091  & 2451793.50 \\
  1.0588 & .0085 & 1.839 & .0090 & 2451799.25 \\
  1.0616 & .0085 & 1.840 & .0090 & 2451807.50 \\
  1.0468 & .0087 & 1.856 &.0102  & 2451815.25 \\
  1.0013 & .0086 & 1.841 &.0094  & 2451834.25 \\
  0.9944 & .0088 & 1.840 & .0107 & 2451869.75 \\
  1.1008 & .0086 & 1.828 &.0094  & 2451907.75 \\
  1.1141 & .0086 & 1.824 & .0085 & 2451913.75 \\
  1.1446 & .0086 & 1.803 & .0091 & 2451922.75 \\
  1.1587 & .0087 & 1.797 & .0095 & 2451927.75 \\
  1.1866 & .0086 & 1.740 & .0090 & 2451950.75 \\
  1.1964 & .0087 & 1.730 &.0093  & 2451958.75 \\
  1.2070 & .0087 & 1.711 & .0093 & 2451966.5 \\
  1.2178 & .0089 & 1.712 & .0099 & 2451975.75 \\
  1.2139 & .0086 & 1.722 &.0089  & 2451983.75 \\
  1.2220 & .0086 & 1.723 & .0089 & 2451984.75 \\
  1.2168 & .0090 & 1.759 & .0101 & 2452007.50 \\
  1.2446 & .0087 & 1.763 & .0093 & 2452007.75 \\
\hline
\end{tabular}
\label{data}
\end{center}
\end{table}

\end{document}